# MAGNETIC COUPLING AND LONG-RANGE ORDER IN THE SPIN-CHAIN SULPHIDE $Ba_2CoS_3$

Andrew D. J. Barnes,[a] Thomas Baikie,[b] Vincent Hardy,[c] Marie B. Lepetit,[c] Antoine Maignan,[c] Nigel A. Young,[a] M. Grazia Francesconi [a]

[a] *Department of Chemistry, University of Hull, Cottingham Road, Hull, HU6 7RX, United Kingdom. Fax: +44 (0)1482 466410 ; Tel: +44 (0)1482 465409; E-mail:* m.g.francesconi@hull.ac.uk
[b] *Nanyang Technological University, Division of Materials Science, 50 Nanyang Avenue, 639798, Singapore. Fax: +65 6790 90818; Tel: +65 6790 6090*
[c] *Laboratoire CRISMAT, UMR CNRS 6508,ENSICAEN, 6 Boulevard du Maréchal Juin, 14050, Caen, Cedex-France. Fax: +33(0)231951600; Tel: 33(0)231452634*

In this paper, we report on the magnetic properties of $Ba_2CoS_3$, a spin-chain compound recently found to be the first $Co^{2+}$ containing one-dimensional sulphide to show metallic-like conductivity and negative magnetoresistance. We carried out an in-depth experimental investigation of the local structure of the cobalt atoms, and ab-initio calculations of the resulting electronic configuration of $Co^{2+}$. From theoretical considerations, the intra-chain coupling was predicted to be antiferromagnetic. Experimentally, several estimates of this magnetic coupling were derived by analysing the temperature dependence of the magnetic susceptibility. Magnetic and heat capacity measurements also provided evidence of a three-dimensional antiferromagnetic ordering, a feature indicative of a noticeable inter-chain coupling in this quasi-1D system.

## 1. Introduction

Traditionally, research on low-dimensional materials is more focussed on oxides, mainly because the preparation and characterisation of non-oxide compounds presents more experimental challenges. Nevertheless, progress in synthetic methods has encouraged the preparation of a wider range of non-oxide materials, such as nitrides or sulphides.[1] It is now becoming apparent that non-oxide materials are an important alternative to traditional oxides to study magnetic interactions in solids in which anions mediate magnetic coupling between transition metal.[2] Low-dimensional magnetic interactions take place among localised unpaired electrons along a chain (one-dimensional structures) or within a plane (two-dimensional structures) and are mediated by the anions bridging the transition metals. The preparation and characterisation of new materials showing low-dimensional magnetism is of paramount importance. The characterisation of one-dimensional materials is particularly welcome, as these systems are rarer than two-dimensional materials.[3]

We have recently concentrated on the $Ba_2CoS_3$ compound, a one-dimensional ternary sulphide, which is the only known inorganic solid containing $Co^{2+}$ in corner-linked [$Co^{2+}$-Anion] tetrahedra forming one-dimensional chains.[4] We reported the synthesis and structural features of $Ba_2CoS_3$ and, subsequently, we found that $Ba_2CoS_3$ shows metallic-like behaviour and negative magnetoresistance, making it the first one-dimensional sulphide containing $Co^{2+}$ to show negative magnetoresistance.[5]

$Ba_2CoS_3$, is structurally similar to $Ba_2MnS_3$[6] and isostructural to other $Ba_2MS_3$ compounds, i.e. $Ba_2ZnS_3$ and $Ba_2FeS_3$.[7] The $Ba_2CoS_3$ and $Ba_2MnS_3$ show the same structural feature of chains of vertex sharing $MS_4$ tetrahedra (M = Co, Mn) and the same one-dimensional antiferromagnetic behaviour. However, there are differences between the structure of $Ba_2CoS_3$ and $Ba_2MnS_3$, which mainly result from differing connectivity among the two crystallographically independent barium ions within the structures of the two sulphides.

$Ba_2MnS_3$ is the best characterised phase within the $Ba_2MS_3$ (M = Mn, Co, Fe, Zn) family of compounds. In 1971, Grey and Steinfink[6] studied the crystal structure and magnetic properties of the sulphide and the isostructural selenide phases and concluded that both materials could be viewed as one-dimensional antiferromagnets. In addition, the authors showed that the magnetic properties of both phases could be modelled using the Heisenberg model developed by Bonner and Fisher[8] and the 'reduced spin' model developed by Emori et al.[9] Using these models the authors were able to determine the intra-chain coupling between the $Mn^{2+}$ ions via the intervening chalcogenide anion.

Following this work, Greaney et al.[10] studied the entire $Ba_2MnQ_3$ system (Q = S, Se and Te), including mixed chalcogenide solid solutions. In this work, the linear chain model of Bonner and Fisher was applied to the susceptibility data and it was observed that the magnitude of the magnetic coupling constant was inversely proportional to the MnMn distance. This result was consistent with the assumption that the magnetic interactions in these chains occur through a superexchange mechanism.

The $Ba_2FeS_3$ and $Ba_2CoS_3$ phases are less well characterised. Synthesis and antiferromagnetic properties of $Ba_2CoS_3$ were reported by Nakayama et al. for the first time but no structural details were included.[11] In 1972, Hong and Steinfink reported the crystal structures of a number of phases within the Ba-Fe-S(Se) system, among which were the isostructural phases $Ba_2FeS_3$ and $Ba_2FeSe_3$.[12] In addition, the authors stated that they had synthesised $Ba_2CoS_3$ and found it was isostructural to $Ba_2FeS_3$. However, no detailed synthetic or crystallographic data were given except for lattice parameters.

We confirmed that $Ba_2CoS_3$ is isostructural with $Ba_2FeS_3$[4] as well as $Ba_2ZnS_3$,[7, 13] $K_2CuCl_3$, , $Cs_2AgCl_3$, $Cs_2AgI_3$, $(NH_4)_2CuBr_3$ and $CuPbBiS_3$.[14] The unit cell is orthorhombic and the cell parameters are $a$ = 12.000(1) Å, $b$ = 12.470(1) Å and $c$ = 4.205(2) Å. The $Ba^{2+}$ cations occupy two distinct crystallographic sites, both coordinated by $S^{2-}$ in a prismatic fashion. One $Ba^{2+}$ is surrounded by six $S^{2-}$ ions at corners of a trigonal prism and an additional $S^{2-}$ is approximately centred above one rectangular face. Seven $S^{2-}$ ions surround the other $Ba^{2+}$ forming a distorted trigonal prism with one face capped. $Co^{2+}$ is tetrahedrally coordinated by $S^{2-}$ and neighbouring Co-S tetrahedra are connected via corners, forming infinite chains. The Co-S tetrahedra are distorted, as the two Co-S bridging bonds are stretched along the chain direction. Those bonds are 2.427(2) Å, which is significantly longer than the two terminal Co-S bonds, 2.330(3) Å and 2.317(3) Å. The bonding angles of S-Co-S deviate from 109.5° by up to several degrees. A similar tetrahedral distortion is also observed in $Ba_2ZnS_3$.[7, 13]

The distance between $Co^{2+}$ cations within each chain (intra-chain) is 4.205(1) Å, whereas the distances between $Co^{2+}$ cations in two neighbouring chains (inter-chain) are 6.153(3) Å and 6.582(3) Å. Neighbouring chains of Co-S tetrahedra are inter-layered by Ba-S blocks. This confers one-dimensional character to the Co-S chains (Figure 1).[4]

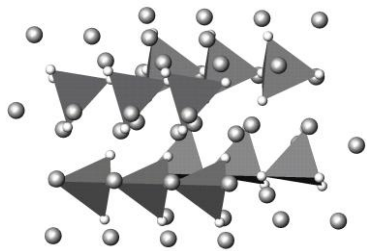

**Fig. 1** Structural representation of $Ba_2CoS_3$: Large dark and small light spheres represent the $Ba^{2+}$ and $S^{2-}$ ions respectively and the grey polyhedra represent the $CoS_4$ tetrahedra.

In previous studies, the plot of the susceptibility versus temperature showed a broad maximum, which is indicative of one-dimensional magnetic ordering and the negative value of θ suggested antiferromagnetic coupling between $Co^{2+}$ cations. [4, 11]

In the present paper, we report numerical values of the coupling constant J calculated using a variety of models. Our results point towards $Ba_2CoS_3$ being a Heisenberg system rather than an Ising system. We also report evidence for a three-dimensional, long-range antiferromagnetic order (LRO) at a relatively high temperature (T = 46 K). We also report EXAFS and XANES spectra, which confirm the distorted tetrahedral coordination of cobalt in $Ba_2CoS_3$ and its +2 oxidation state, and exact calculations on the $Co^{2+}$ magnetic state, showing that in this distorted tetrahedral coordination, $Co^{2+}$ has a S=3/2 high spin ground state. This result constitutes the basis for calculations on the microscopic nature of the antiferromagnetic coupling between $Co^{2+}$ mediated by $S^{2-}$.

## 2. Experimental

The synthesis of $Ba_2CoS_3$ was achieved via a solid-gas reaction between stoichiometric mixtures of $BaCO_3$ and cobalt metal powder or cobalt oxides/carbonates ($CoCO_3$, $CoO$, $Co_3O_4$) with $CS_2$ vapour. Mixtures of starting reagents were ground and loaded into carbon boats. The reactions were carried out in a tubular furnace under a $CS_2/N_2$ flow at temperatures ranging 800-1200 ºC for 12-48 hours. Once the reactions were complete, the samples were slow cooled to room temperature.

The Co K-edge XAS were collected in transmission mode at *ca.* 80 K from samples diluted in boron nitride on station 9.2 of the Daresbury Laboratory SRS (2 GeV 100-200 mA) using Si(111) or Si(220) double crystal monochromators with 50% harmonic rejection. The spectra were averaged and calibrated (first maximum of the first derivative of the Co-K edge of Co foil (7709.0 eV)) using PAXAS.[15] The data were modelled using curved wave theory in EXCURV 98.[16] Co(salen) [bis(µ$_2$-(bis(salicylaldehyde)-ethylenedi-iminato-N,N',O,O,O')-cobalt(II))], [Co(acac)$_2$.H$_2$O]$_2$ [bis(aqua-( µ$_2$-acetylacetonato-O,O,O')-(acetylacetonato)-cobalt(II))] and CoBr$_2$(PPh$_3$)$_2$ [dibromo-bis(triphenylphosphine)cobalt(II)] were obtained from Aldrich and used as supplied. Co3,5-$^t$Bu-cysalen [((R,R)-(-)-N,N'-bis(3,5-Di-t-butylsalicylidene)-1,2- cyclohexanediamino)cobalt(II)] was a kind gift from Prof. Woodward (University of Nottingham).[17]

Magnetic measurements were carried out using a SQUID magnetometer (Quantum Design) over the temperature range 5 - 400 K. Measurements of χ (T) were recorded in a magnetic field of 1 T, i.e. a quite large value was needed to enhance the signal-to-noise ratio. This value, though, was checked to be within the linear regime of M (H) over the whole T-range. Sample mounting, sample size and the subtraction procedure of the background signal were also optimised in order to maximise the quality of the data. Heat capacity measurements were recorded by means of a PPMS (Quantum Design), using a relaxation method with a 2τ fitting procedure.

## 3. Results and Discussion

### 3.1 Spectroscopic characterisations of the cobalt sites

**3.1.1 Co K-Edge XANES.** Co K-edge XANES for $Ba_2CoS_3$ as well as those for some representative $Co^{2+}$ model compounds are shown in

**Fig. 2** Co K-edge XANES spectra of (a) $[Co(acac)_2.H_2O]_2$, (b) $[Co(salen)]_2$, (c) Co3,5-$^t$Bu-cysalen (d) $CoBr_2(PPh_3)_2$ and (e) $Ba_2CoS_3$.

**Fig. 3** Co K-edge EXAFS (a) and Fourier Transform (b) for $Ba_2CoS_3$.

. The edge position shifts by about 3 eV to higher energy for $Co^{3+}$ compounds. Both the weak features just before the edge, which are associated with 1s-3d transitions, as well as the features on the edge, which can be thought of as 1s-4p transitions, can be used to identify the site symmetry of the Co environment. The intensity of the 1s-3d pre-edge features is very dependent on the symmetry of the Co environment, and in particular the presence of a centre of symmetry and/or extensive p-d mixing of the 3d orbitals. Therefore in in octahedral $(Co(acac)_2.H_2O$[18] (Figure 2(a)) and square-planar[17,19] (Co3,5-$^t$Bu-Cysalen (Figure 2(c)) geometries the 1s-3d transitions are very weak, but in the spectra of square based pyramidal[20] (Co(salen) (Figure 2(b)) and tetrahedral[21] $CoBr_2(PPh_3)_2$ (Figure 2(d)) environments the 1s-3d transitions have greater intensity. The presence of well-defined shoulders on the edge itself have been assigned previously to 1s-4$p_z$ transitions (with shakedown contributions) in tetragonal geometries where one or more axial ligands are absent.[22, 23] For square-planar Co3,5-$^t$Bu-Cysalen (Figure 2(c)), this shoulder is more marked than for square-based pyramidal Co(salen) (Figure 2(b)). Therefore, the relatively intense pre-edge feature at 7708.0 eV in $Ba_2CoS_3$ is consistent with a tetrahedral environment for the Co, but not an octahedral or square-planar geometry. The lack of a shoulder on the edge also rules out a square-based pyramidal geometry. The overall structure of the edge features are also very similar to those observed for tetrahedral $CoBr_2(PPh_3)_2$. Although the 1s-3d pre-edge feature in $Ba_2CoS_3$ is less well resolved and a little lower in energy than the pre-edge feature in $CoBr_2(PPh_3)_2$ at 7708.6 eV, this behaviour is similar to that observed on increasing the number of sulphur ligating atoms in $Ni^{2+}$ complexes,23 as is the observation that the edge is also at a slightly lower energy edge position for $Ba_2CoS_3$ than $CoBr_2(PPh_3)_2$. Therefore, the Co K-edge XANES of $Ba_2CoS_3$ is completely consistent with a distorted tetrahedral $CoS_4$ environment for the Co.

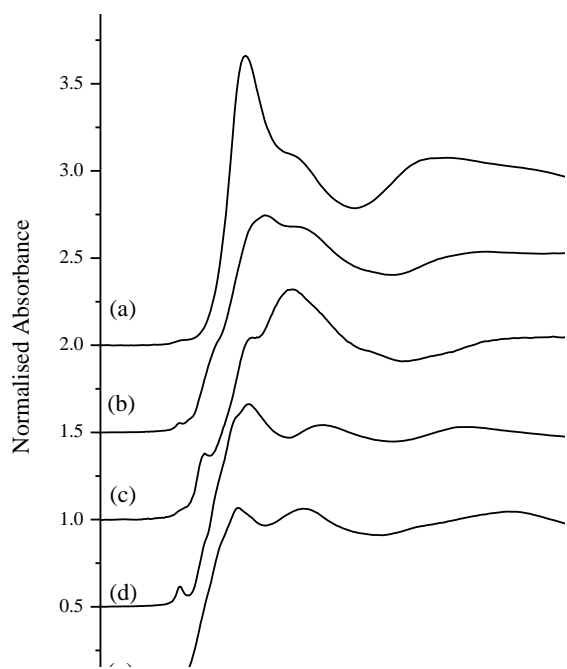

**Fig. 2** Co K-edge XANES spectra of (a) [Co(acac)$_2$.H$_2$O]$_2$, (b) [Co(salen)]$_2$, (c) Co3,5-$^t$Bu-cysalen (d) CoBr$_2$(PPh$_3$)$_2$ and (e) Ba$_2$CoS$_3$.

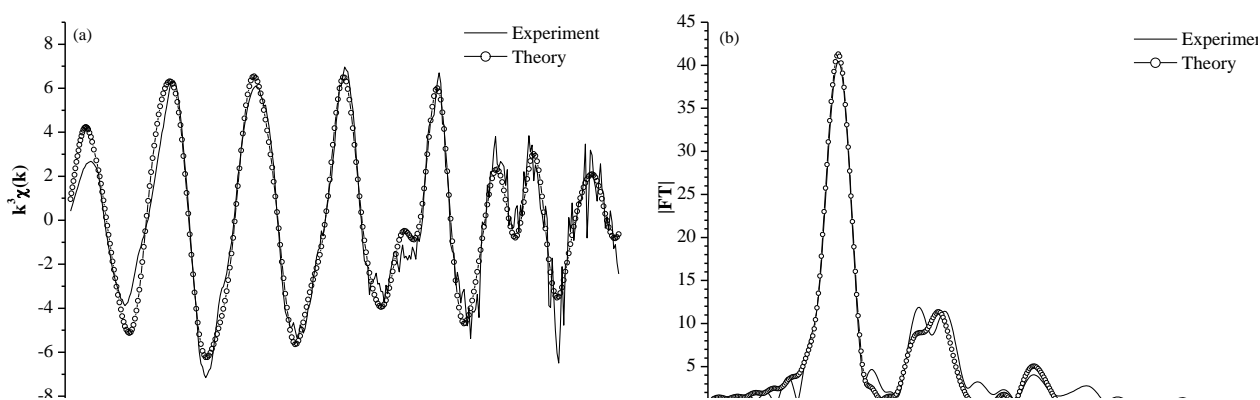

**Fig. 3** Co K-edge EXAFS (a) and Fourier Transform (b) for Ba$_2$CoS$_3$.

**3.1.2 Co K-edge EXAFS.** The Co K-edge EXAFS and Fourier Transforms (FT) are shown in **Erreur ! Source du renvoi introuvable.**respectively and the structural parameters obtained from the fitting of the data are listed in **Erreur ! Source du renvoi introuvable.**. The local environment of the Co was initially set using a combination of the XRD data for Ba$_2$CoS$_3$ and the literature data for the isostructural Ba$_2$ZnS$_3$.13 The Co-S shell is formally split into 2.328, 2.318, 2.427 and 2.427 Å components. The initial attempts to refine these separately as two shells (2.33 and 2.43 Å) were unsuccessful, with the refinement merging to give essentially identical Co-S bond lengths of 2.340 Å, but with relatively large Debye-Waller factors. Therefore, for all subsequent refinements the Co-S shell was taken as a single shell with occupation number 4 and this was confirmed by mapping the occupation number versus the Debye-Waller factor for this shell. The large Debye - Waller factor is indicative of a distorted environment.

| Bond | $r$/Å | $2\sigma^2$/Å$^2$ |
|---|---|---|
| **Co-S (x4)** | 2.341(4) | 0.0149(6) |
| **Co-Ba (x3)** | 3.553(11) | 0.0224(16) |
| **Co-Ba (x4)** | 3.873(17) | 0.0224(16) |
| **Co-Co (x2)** | 4.090(12) | 0.0119(39) |
| **Co-Ba (x4)** | 5.554(15) | 0.0170(29) |

| $E_f$ = -8.6(5) | R = 26.12 % | FI = 0.30 |
|---|---|---|

**Table 1** Summary of refined Co K-edge EXAFS parameters for $Ba_2CoS_3$.

The next significant shell in the FT occurs as a complex feature between 3 and 5 Å. This is made up of a mixture of Co-Ba and Co-Co interactions. Whilst the Co-Co has an occupation number of 2, the 7 Co-Ba interactions are more disordered with a range of inter-atomic distances (3.515 Å (×2), 3.693 Å (×1), 3.855 Å (×2) and 3.856 Å (×2)). Fitting this many shells to a relatively weak feature is a challenge for EXAFS and this was partially overcome by setting the Debye-Waller for the two refined Co-Ba shells at 3.55 and 3.87 Å to be equal to each other . The fact that these refined Co-Ba Debye-Waller factors are greater than the Co-Co Debye-Waller factor is understandable, as the $CoS_4$ tetrahedra are vertex sharing, whereas the Co-Ba interactions are between the layers, but too much should not be read into the numbers. The final shell in the FT to be fitted was that at about 5.5 Å and this is another Co-Ba shell, which in this case easily fitted to four Co-Ba distances with a reasonable Debye-Waller factor. Therefore, the agreement between the Co K-edge EXAFS data and the PXRD data out to at least 5.5 Å confirms the structure.

### 3.2   Theoretical predictions on the magnetism of $Ba_2CoS_3$

**3.2.1 Ab-initio calculations on the magnetic state of $Co^{2+}$.** EXAFS spectroscopy confirms that, in $Ba_2CoS_3$, the $Co^{2+}$ ion is in a distorted tetrahedral environment. In a perfect tetrahedral environment, corresponding to the local $T_d$ point group, the $Co^{2+}$ ion would see its *3d* atomic orbitals split into two *E* orbitals ($z^2 - r^2$ and $x^2 - y^2$) and three $T_2$ orbitals ($xz$, $yz$, $xy$). In the present system, the tetrahedron is not only elongated along the crystallographic *c* direction, which would yield a $C_{2v}$ local point group, but further distorted in a $C_s$ subgroup. Therefore, the only remaining symmetry operation is the plane orthogonal to the *c* direction and containing the cobalt atom. This crystal field splits the *3d* orbitals further, lifts all remaining degeneracies and gives three non degenerate *A′* orbitals and two non degenerate *A″* orbitals.

In addition to this crystal field splitting, which is a single electron effect, the strong correlation effects occurring in the *3d* shell of the first-row transition-metal based compounds re-normalise the splitting between the *3d* orbitals further. In order to evaluate accurately the effective splitting of the *3d* orbitals of the $Co^{2+}$ ion, we performed ab-initio calculations of the low energy states for $Co^{2+}$ ion in a $C_s$ symmetry. We computed the $Co3d \rightarrow Co3d$ electronic excitations of a fragment of the crystal, embedded in a field of point charges and pseudo potentials, reproducing the main effects of the rest of the crystal on the studied fragment. The fragment is composed of one cobalt atom and its four sulphur ligands. The embedding is designed to reproduce the correct Madelung potential and the exclusion effects of the electrons of the rest of the crystal on the fragment.[24] The method used to calculate the ground and excited states treats the correlation effects of the cobalt *3d* shell, within a complete active space self-consistent calculation[25] as well as the screening effects due to the other electrons of the cobalt and those of the sulphide anions in the first coordination shell. The screening effects have been shown to decrease as $1/R^4$ and $1/R^6$ as a function of the distance of the screening excitation to the cobalt *3d* shell and can be considered to be correctly reproduced within such a first shell approximation.[26,27] The method used for the spectroscopic calculations is the Difference Dedicated Configuration Interaction,[28] plus Davidson correction. This method was specifically derived for accurate treatment of the electronic spectroscopy of open-shell systems and proved its efficiency for the accurate determination of local quantities in strongly correlated systems.[29]

The calculated orbitals for the $Co^{2+}$ ion are shown in Figures 4(a), 4(b) and 4(c) (*A'*) and in 4(d) and 4(e) (*A''*). The local quantification axis for $Co^{2+}$ does not correspond to the c axis but is now defined as the axis passing through the $Co^{2+}$ centre and the midpoint between the two sulphur anions in the *c* chains direction: ($S_3$, $S_4$). This axis is within the $C_s$ symmetry plane, while the two other axes can be defined as the *x* direction coinciding with the *c* crystallographic axis ($S_3$, $S_4$) and the *y* axis being the orthogonal axis to the preceding ones.

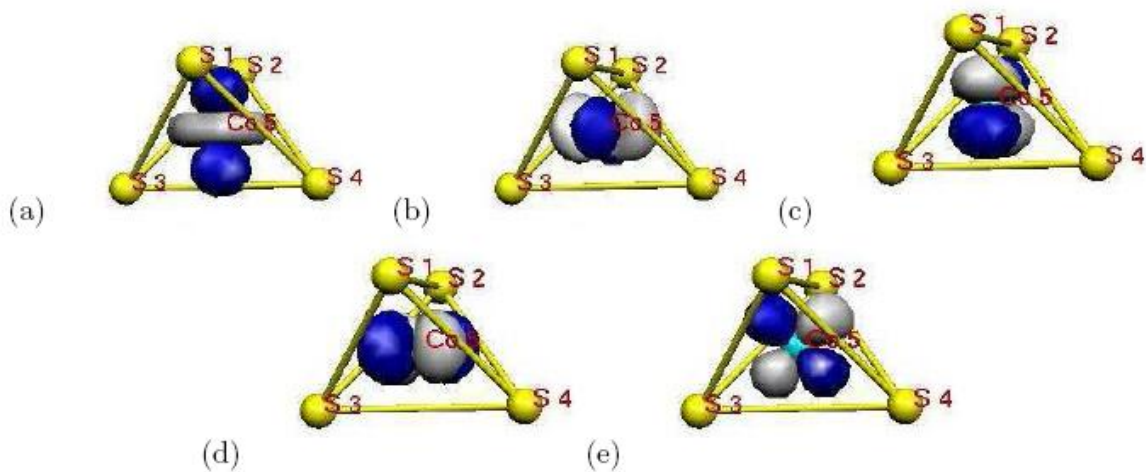

**Fig. 4** *3d* atomic orbitals of the $Co^{2+}$ ion in the $Ba_2CoS_3$ environment. The crystallographic *c* axis is along the $(S_3, S_4)$ direction. The reflection plane is perpendicular to the $(S_3, S_4)$ direction, passing through the Co atom and the $(S_3, S_4)$ midpoint. This plane contains the $Co^{2+}$ ion.

In this new set of axes, the orbitals on $Co^{2+}$ can be labelled and the ground state can be seen in a single Slater's determinant approximation, as shown in Fig. 5.

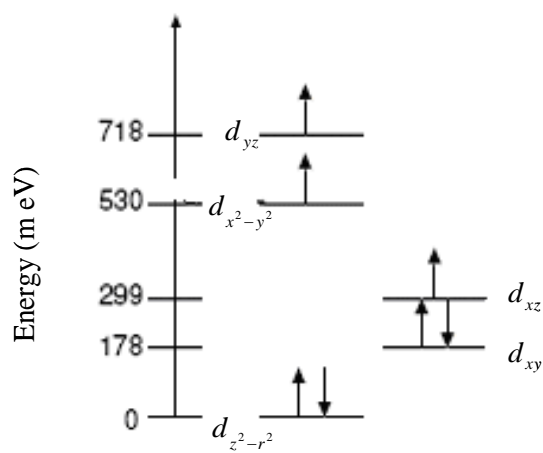

**Fig. 5** Orbital diagram of the $Co^{2+}$ ground state.

It is clear from Fig. 5 that the tetrahedral degeneracy of the *3d* orbitals is lifted and that the resulting $Co^{2+}$ orbitals splitting is quite large. The ground state is a high spin, S = 3/2 state with L = 0. Table 2 gives the first excited states of the fragment and the nature of the corresponding electronic states. All states are high spin (S = 3/2), the first excited low spin (S = 1/2) state being at more than 2 eV above the ground state.

| Sym. | $\Delta E$ (meV) | Wave function dominant term |
|---|---|---|
| A' | 148.50 | $+.82|d_{z^2-r^2}d_{x^2-y^2}d_{yz}d_{xy}^2 d_{xz}^2\rangle + .47|d_{z^2-r^2}^2 d_{x^2-y^2}d_{yz}^2 d_{xy}d_{xz}\rangle$ |
| A' | 199.93 | $+.95|d_{z^2-r^2}^2 d_{x^2-y^2}^2 d_{yz}d_{xy}d_{xz}\rangle$ |
| A' | 477.77 | $+.51|d_{z^2-r^2}^2 d_{x^2-y^2}d_{yz}^2 d_{xy}d_{xz}\rangle + .73|d_{z^2-r^2}d_{x^2-y^2}^2 d_{yz}^2 d_{xy}d_{xz}\rangle + .32|d_{z^2-r^2}d_{x^2-y^2}^2 d_{yz}d_{xy}^2 d_{xz}^2\rangle$ |
| A' | 1413.25 | $-.62|d_{z^2-r^2}^2 d_{x^2-y^2}d_{yz}^2 d_{xy}d_{xz}\rangle + .59|d_{z^2-r^2}d_{x^2-y^2}^2 d_{yz}^2 d_{xy}d_{xz}\rangle + .34|d_{z^2-r^2}d_{x^2-y^2}d_{yz}d_{xy}^2 d_{xz}^2\rangle$ |
| A" | 0.00 | $+.68|d_{z^2-r^2}^2 d_{x^2-y^2}d_{yz}d_{xy}^2 d_{xz}\rangle$ |
| A" | 225.25 | $+.68|d_{z^2-r^2}^2 d_{x^2-y^2}d_{yz}d_{xy}d_{xz}^2\rangle - .60|d_{z^2-r^2}d_{x^2-y^2}^2 d_{yz}d_{xy}^2 d_{xz}\rangle - .28|d_{z^2-r^2}d_{x^2-y^2}^2 d_{yz}d_{xy}d_{xz}^2\rangle$ |
| A" | 369.39 | $+.38|d_{z^2-r^2}^2 d_{x^2-y^2}d_{yz}d_{xy}^2 d_{xz}\rangle - .45|d_{z^2-r^2}d_{x^2-y^2}^2 d_{yz}d_{xy}d_{xz}^2\rangle$ |
|  |  | $+.25|d_{z^2-r^2}d_{x^2-y^2}d_{yz}^2 d_{xy}^2 d_{xz}\rangle + .70|d_{z^2-r^2}d_{x^2-y^2}^2 d_{yz}d_{xy}d_{xz}^2\rangle$ |
| A" | 407.12 | $+.64|d_{z^2-r^2}d_{x^2-y^2}^2 d_{yz}d_{xy}d_{xz}^2\rangle - .45|d_{z^2-r^2}d_{x^2-y^2}d_{yz}^2 d_{xy}^2 d_{xz}\rangle + .49|d_{z^2-r^2}d_{x^2-y^2}^2 d_{yz}^2 d_{xy}d_{xz}^2\rangle$ |
| A" | 1150.44 | $+.63|d_{z^2-r^2}d_{x^2-y^2}^2 d_{yz}d_{xy}d_{xz}^2\rangle + .44|d_{z^2-r^2}d_{x^2-y^2}^2 d_{yz}d_{xy}d_{xz}\rangle + .52|d_{z^2-r^2}d_{x^2-y^2}d_{yz}^2 d_{xy}d_{xz}\rangle$ |
| A" | 1460.72 | $+.83|d_{z^2-r^2}d_{x^2-y^2}^2 d_{yz}d_{xy}d_{xz}\rangle + .40|d_{z^2-r^2}d_{x^2-y^2}d_{yz}^2 d_{xy}d_{xz}^2\rangle$ |

**Table 2**: Excitation energy of the $Co^{2+}$ 3d 咯 3d excitations and related wave functions.

**3.2.2 Microscopic origin of the $Co^{2+}$ - $Co^{2+}$ antiferromagnetic coupling.** $Ba_2CoS_3$ shows one-dimensional antiferromagnetism within the Co-S chains. We used the results from the ab-initio calculations to understand the microscopic reason of the antiferromagnetic character of the coupling constant. We analysed the coupling J between two $Co^{2+}$ nearest neighbour ions using a perturbative approach. Importantly, the first non-negligible term comes at the fourth order of perturbation and can be associated with the classical spin flip mechanism mediated by a ligand orbital (see Fig. 6).

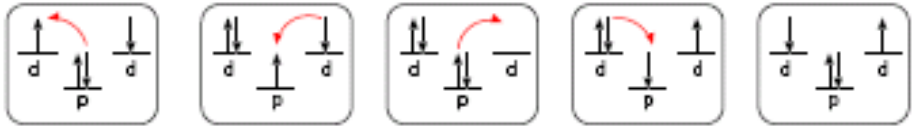

**Fig. 6** Schematic spin flip mechanism between two d orbitals mediated by a ligand *p* orbital.

Indeed, the terms involving direct exchange integrals or direct transfer integrals between *3d* orbitals of the two $Co^{2+}$ ions can be neglected in comparison with the terms mediated by the orbitals of the sulphur atom linking the two tetrahedra, as the direct transfer integrals and the direct exchange integrals decrease exponentially with the $Co^{2+}$ $Co^{2+}$ inter-atomic distance. Fig. 7 shows the cobalt and sulphur orbitals involved in the exchange mechanism. An analysis of the overlap between the *3d* orbitals of the two $Co^{2+}$ ions and the *3p* orbitals of the bridging $S^{2-}$ shows that four different terms can flip spins between the two $Co^{2+}$ ions at the first non-negligible order of perturbation.

1. The spin flip between the two $3d_{x^2-y^2}$ orbitals is mediated by both the sulphur $3p_x$ and $3p_z$ orbitals (see Figure 7). The resulting coupling term is

$$J_{x^2-y^2} = -2 \frac{\left(t^2_{x^2-y^2,z} - t^2_{x^2-y^2,x}\right)^2}{\Delta^2 U} + \text{small terms}$$

where the small terms involve double charge transfer from the ligand to the metal atoms. U is the on-site Coulomb repulsion between two electrons in the same cobalt $3d$ orbitals, $t_{x^2-y^2,x}$ is the transfer integral between the $3d_{x^2-y^2}$ $Co^{2+}$ orbitals and the $3p_x$ bridging sulphur orbital. Likewise, $t_{x^2-y^2,z}$ is the transfer integral between the $3d_{x^2-y^2}$ $Co^{2+}$ orbitals and the $3p_z$ bridging sulphur orbital. $\Delta$ is the energy of the single charge transfer configuration. In this process, one electron from one of the $3p$ sulphur orbitals has been transferred to one of the cobalt $3d$ orbitals and $\Delta$ should depend on the sulphur and cobalt orbitals as well as on the cobalt oxidation state. However, in a first approximation, we will assume that all the single charge transfer configurations are at the same energy $\Delta$.

2. The spin flip between the two $3d_{yz}$ orbitals is mediated by the $3p_y$ orbital on sulphur (see Fig. 7). This interaction results in a coupling term

$$J_{yz} = -4 \frac{t^4_{yz,y}}{\Delta^2 U} + \text{small terms}$$

where $t_{yz,y}$ is the transfer integral between the $3d_{yz}$ orbitals and the $3p_y$ sulphur orbital.

3. The spin flip between the two $3d_{xz}$ orbitals is mediated by both the sulphur $3p_x$ and $3p_z$ orbitals (see Fig. 7). It results in a coupling term

$$J_{xz} = -2 \frac{\left(t^2_{xz,x} - t^2_{xz,z}\right)^2}{\Delta^2 U} + \text{small terms}$$

where the $t_{xz,x}$ is the transfer integral between the $3d_{xz}$ orbitals and the $3p_x$ sulphur orbital. Likewise, $t_{xz,z}$ is the transfer integral between the $3d_{xz}$ orbitals and the $3p_z$ sulphur orbital.

4. Finally there is a crossed term that flips the spin of the $3d_{x^2-y^2}$ orbital of one cobalt cation with the $3d_{xz}$ orbital of the other cobalt. The resulting coupling term can be expressed as

$$J_c = -2\frac{\left(t_{x^2-y^2,z}\, t_{xz,z} - t_{x^2-y^2,x}\, t_{xz,x}\right)^2}{\Delta^2\, U} + \text{small terms}$$

The minus sign in the numerator derives from the fact that the transfer integrals between the S $3p_x$ and Co $3d_{x^2-y^2}$ orbitals, and the S $3p_z$ and Co $3d_{xz}$ orbitals change sign according to which cobalt cation an electron is transferred to or from (see Fig. 7). All exchange terms derived from the above microscopic analysis correspond to only one spin flip at the time.

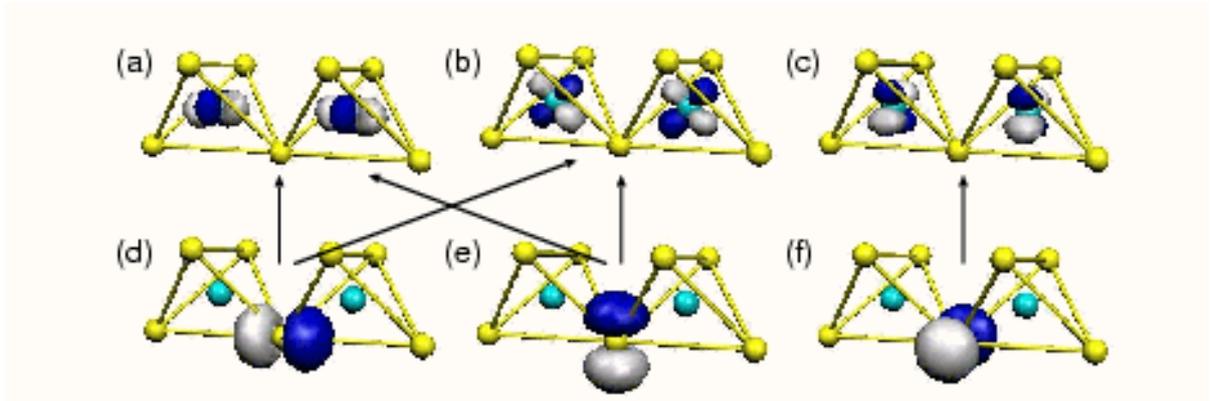

**Fig. 7** Schematic representation of $Co^{2+}$ magnetic orbitals on two neighbouring atoms and $S^{2-}$ orbitals bridging the interactions between the preceding ones. The arrows picture the large overlaps between the magnetic and bridging orbitals mediating the spins exchanges. (a) $Co^{2+}$ $3d_{x^2-y^2}$ (b) $Co^{2+}$ $3d_{xz}$ (c) $Co^{2+}$ $3d_{yz}$ magnetic orbitals (d) $S^{2-}$ $3p_x$; (e) $S^{2-}$ $3p_z$; (f) $S^{2-}$ $3p_y$ bridging orbitals. The $3d_{x^2-y^2}$ and $3d_{xz}$ orbitals of the $Co^{2+}$ cations overlap with both the $3p_x$ and $3p_z$ orbitals of the bridging $S^{2-}$ anion. The sign of the transfer integral between the $3d_{x^2-y^2}$ and $3p_x$ orbitals, and the $3d_{xz}$ and $3p_z$ orbitals is different for the two cobalt cations, while this is not the case for the the other cobalt-sulphur overlaps. This sign change is responsible for the minus sign appearing in numerator of the $J_{x^2-y^2}$, $J_{xz}$ and $J_c$ exchange integrals.

In order to determine whether these exchange terms result in a global antiferromagnetic interaction the interaction matrix between two cobalt cations needs to be diagonalised. One can work without loss of generality in the total $S_z = 0$ sector. The interaction matrix between two $Co^{2+}$ ions thus involves all possible ways to put three spins up and three spins down in six singly-occupied orbitals (three on each $Co^{2+}$ ions). It results in a 20 × 20 matrix (see appendix), which can be analytically diagonalised. All coupling terms $J_{x^2-y^2}$, $J_{yz}$, $J_{xz}$ and $J_c$ are negative and it is possible to determine the ground state, whatever the numerical values of the basic exchange integrals. If the $S = 3$ state, associated with the global ferromagnetic interaction between two cobalt cations, is taken as the zero of energy, the ground state energy can be expressed as

$$E_{GS} = J_C \vee J_{yz} \vee \frac{J_{xz}}{2} \vee \frac{J_{x^2-y^2}}{2} - \frac{1}{2}\sqrt{(J_{x^2-y^2} \vee J_{xz} - J_C)^2 + 3J_C^2} \quad (1)$$

The ground state wave function is the singlet combination of high spin $S = 3/2$ atomic states (see appendix).

We have thus shown that, whatever the amplitudes of the basic exchange terms, the global interaction between the $Co^{2+}$ ions is antiferromagnetic.

### 3.3 Experimental determination of the intra-chain coupling $J$

In this section, we determine values of the intra-chain antiferromagnetic coupling $J$, i.e. the coupling constant between two $Co^{2+}$ cations bridged by one $S^{2-}$ anion within the same one-dimensional chain, by performing different analyses of the susceptibility curve $\chi(T)$.

It must be noted that detailed analysis of the magnetism associated with $Co^{2+}$ cations have been carried out in the case of molecular materials.[30, 31] To our knowledge, however, similar models of the magnetic susceptibility are not available in the case of infinite chains of $Co^{2+}$ spins in distorted tetrahedral environment

Fig. 8 shows the temperature dependence of the magnetic susceptibility for $Ba_2CoS_3$ (main panel) along with the reciprocal curve (inset).

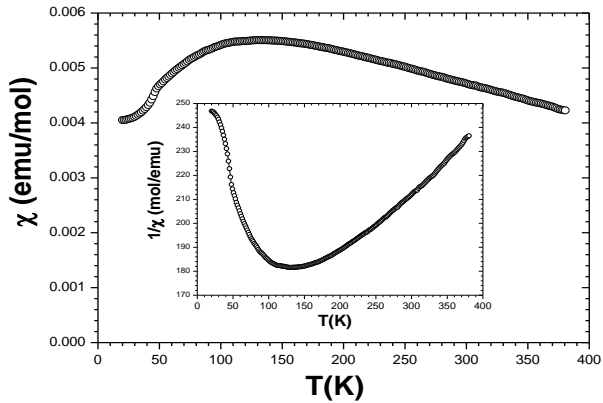

**Fig. 8** Main panel : Temperature dependence of the magnetic susceptibility $\chi$. Inset: Temperature dependence of the reciprocal susceptibility $1/\chi$.

There is excellent agreement between this set of measurements and the data reported earlier,[4,11] specifically in the global shape of the curve, the mean value of $\chi$ (~0.005 5 emu/mol), the coordinates of the smooth maximum ($\chi_{max}$ and $T_{max}$), as well as the presence of an upturn at a temperature below 20 K. Such an upturn has been regarded as a "Curie tail" related to the unavoidable presence of a small amount of impurities and/or to broken chains.4

In this paper, we present only the data for T > 20 K. As previously discussed, the smooth bump in χ (T) can be ascribed to the typical behaviour expected for one-dimensional antiferromagnetic chains. Qualitatively, such a broad maximum in χ (T) results from the competition between pure paramagnetism (as expected for isolated spins) and antiferromagnetic fluctuations along each chain, associated with intra-chain spin coupling (*J*). In a first step, let us focus on the high-T range, i.e. for T >> $T_{max}$.

### 3.3.1 High-T range of the χ (T) curve.

**3.3.1.1 Curie-Weiss approach.** The simplest approach consists in focussing on the Curie-Weiss (CW) regime, which should take place for $T >> |J|$. Within the CW regime, the χ (T) dependence is expressed by:

$$\chi = \frac{C}{T - \theta_{CW}} \qquad (2)$$

where C is the Curie constant and $\theta_{CW}$ is the Curie-Weiss temperature, expressed by

$$C = \frac{N_A g^2 \mu_B^2 S(S+1)}{3k_B} \quad \text{and} \quad \theta_{CW} = \frac{2zJS(S+1)}{3} \qquad (3)$$

where *J* is in Kelvin, *z* is the number of nearest-neighbours involved in the exchange coupling (z = 2 for intra-chain coupling), S is the spin value (3/2 in our case), $N_A$ is the Avogadro number, g the Landé factor, $\mu_B$ the Bohr magneton and $k_B$ the Boltzman constant. The above equation for $\theta_{CW}$ corresponds to an exchange term in the Hamiltonian of the form *(-2J $S_i S_j$)* with $i \neq j$ (*J* < 0 for antiferromagnetic coupling).

The values of C and $\theta_{CW}$ allow to derive the effective magnetic moment $\mu_{eff} = p_{eff} \mu_B \simeq \sqrt{8C} \mu_B$ (with C in emu·K·mol$^{-1}$) or the Landé factor $g = p_{eff}/\sqrt{S(S+1)}$, as well as the intra-chain coupling $J = 3\theta_{CW}/4S(S+1)$.

A linear regime for $1/\chi$ versus T at high temperatures is required in order to apply the CW model to calculate *J*. However, it turns out that, for Ba$_2$CoS$_3$, a true linear regime is never reached over the investigated T-range.

Close inspection of the curve shown in the inset of Fig. 8 reveals that a curvature is still present up to 380 K. Such a feature was already pointed out by Nakayama *et al.* in their investigation of Ba$_2$CoS$_3$.[11] The use of a linear fitting over a limited T range at the highest temperatures is often used with the aim of deriving values of $\theta_{CW}$ and $p_{eff}$ (i.e. g and *J*), but it may be misleading. For instance, a linear fitting performed on the curve of the inset of Fig. 8, in the range 300-230 K would give $\theta_{CW} \approx$ -546 K and $p_{eff} \approx$ 5.65, while a linear fitting in the range 380-300 K, would give $\theta_{CW}$ = -381 K and $p_{eff}$ ≈ 5.08. We also emphasise that the $p_{eff}$ obtained above are suspiciously larger than the usually observed values for $Co^{2+}$ ($p_{eff}$ ≈ 4.8). The above considerations clearly indicate a failure of the simple CW model to fit the $1/\chi$ (T) curve for $Ba_2CoS_3$, at the highest temperatures. Therefore, in order to calculate the constant $J$, more sophisticated models are necessary, which all depend on the degree of magnetic anisotropy. In other words, one has to know whether $Ba_2CoS_3$ should be treated either in the Heisenberg-like or in the Ising-like limit.

**3.3.1.2 Magnetic anisotropy**. As in most compounds, the main source of possible anisotropy in $Ba_2CoS_3$ should originate from the crystalline electric field (CEF). It was discussed above that $Co^{2+}$ is in a distorted tetrahedral environment of four $S^{2-}$ anions nearest-neighbours, modelled by the point group $C_s$.

In this situation, it must be emphasised that the spin-orbit (SO) coupling has to be taken into account to address properly the magnetic anisotropy. This is not an easy task and we emphasise that SO was omitted in the ab-initio calculations reported here in section 3.2.2. Instead, let us consider a simpler and more standard approach, in which all the interaction terms are taken into account via successive perturbations. Accordingly, the energy level diagram for a single $Co^{2+}$ can be derived by considering firstly the tetrahedral part of the CEF and secondly both the SO interaction and the departure from the $T_d$ group.[32] For $Co^{2+}$ which has a $3d^7$ configuration, the $T_d$ part of the CEF leads to a ground state $^4A_2$, which is a quartet with L = 0 and S = 3/2.[31] However, there is a certain amount of admixture from higher-lying states with L ≠ 0 associated with second order effects of CEF. Therefore, the combination of spin-orbit and axial distortion lifts the degeneracy of the S=3/2 ground state. Since $Co^{2+}$ is a Kramers ion, this splitting of the $^4A_2$ manifold leaves two doublets: one corresponding to $S_z$ = ± 1/2 and the other corresponding to $S_z$ = ± 3/2. This situation can be described by the introduction of a $-DS_z^2$ term in the single-site Hamiltonian, with $D$ being the single-ion anisotropy. Note that positive D values favour a spin alignment along z and the energy spacing between the two doublets of $Co^{2+}$ is equal to 2D. Since D is a single-ion feature, the order of magnitude of this parameter in $Ba_2CoS_3$ can be estimated from the values found in compounds that contain $Co^{2+}$ in similar CEF. The $D$ values found in the literature range between 0.1 and 10 K.[33]

In such an energy diagram, one can anticipate that only the lowest doublet is populated at very low-temperature, which must induce a large magnetic anisotropy (i.e. Ising-like limit for $T \ll D$). As the temperature is increased, however, the influence of the splitting between the two doublets progressively vanishes, which tends to reduce the effect of the anisotropy (i.e. Heisenberg-like limit). Without a precise determination of $D$, none of these limits can be *a priori* discarded and one must start the analysis by considering both of them.

**3.3.1.3 Series expansion.** It was shown above that the pure CW regime is not reached in data, however, it may be possible to analyse the high-T magnetic susceptibility by using the formula of series expansion. In this approach, χ is developed in series of powers of $|J|/T \ll 1$:

$$\frac{1}{\chi_r} = \frac{3\theta}{S(S+1)} \sum_{n=0}^{n=\infty} \frac{b_n}{\theta^n} \quad (4)$$

where $\chi_r = \frac{k_B |J| \chi}{N_A g^2 \mu_B^2}$ is the reduced susceptibility, and $\theta = T/|J|$ is the reduced temperature.

A key point to emphasise is that the coefficients $b_n$ of such a series are not approximations but they are derived from exact calculations. Theoretical models were developed for the two limits: in the Heisenberg limit by Rushbrooke and Wood (up to n=6)[34] and in the Ising limit by Emori et al. (up to n=2).[9] In series expansions, the influence of each term decreases rapidly with the value of n. It is thus possible to get a good correction to the pure CW law at high-T by simply considering the first two terms. For $Ba_2CoS_3$, we found that the series expansions associated with both the Heisenberg and Ising models can be written in the following form:

$$\frac{1}{\chi_r} \equiv \frac{3\theta}{S(S+1)} \left\{ 1 - \frac{2zS(S+1)}{3\theta} - \frac{2zS(S+1)}{3\theta^2} \alpha \right\}, \quad (5)$$

where α is a parameter which depends only on the model and on the value of S. In the Heisenberg limit: $\alpha = -(1/3)(S(S+1)/2)$ whereas in the Ising limit: $\alpha = -(1/5)(S(S+1)/5)$. Re-introducing the susceptibility, χ, in place of the reduced susceptibility, $\chi_r$, leads to:

$$\frac{N_A g^2 \mu_B^2}{k_B \chi} \equiv \frac{3T}{S(S+1)} - 2z|J| \left[ 1 - \frac{\alpha |J|}{T} \right] \quad (6)$$

Finally, considering that z = 2 along a chain and using practical cgs units, the above equation can be rewritten as:

$$F_g[\chi, T] = G_J[T] \quad (7)$$

where $F_g$ and $G_J$ are two functions defined as follows:

$$F_g[\chi(T),T] = 0.09378\frac{g^2}{\chi} - \frac{3}{4S(S+1)}T$$

$$G_J[T] = -J + \frac{\alpha J^2}{T}$$

In these functions, the parameters to be adjusted are g and *J*. The only criterion is to find pairs of values that fit Equation 7 in the highest-T range. It must be emphasised that α is not an adjustable parameter. Rather, its value is determined by the type of exchange considered: α = 3 in the Heisenberg case, or α = (17/10) in the Ising case.

$F_g[\chi(T),T]$ derives from the experimental $\chi(T)$ curve, once a value of the Landé factor g has been chosen. $G_J[T]$ depends only on *J*, once the value of α has been fixed. The search for appropriate pairs of values for g and *J* is performed by trial and error.

A systematic search has been performed and selected plots are displayed in Figure 9.

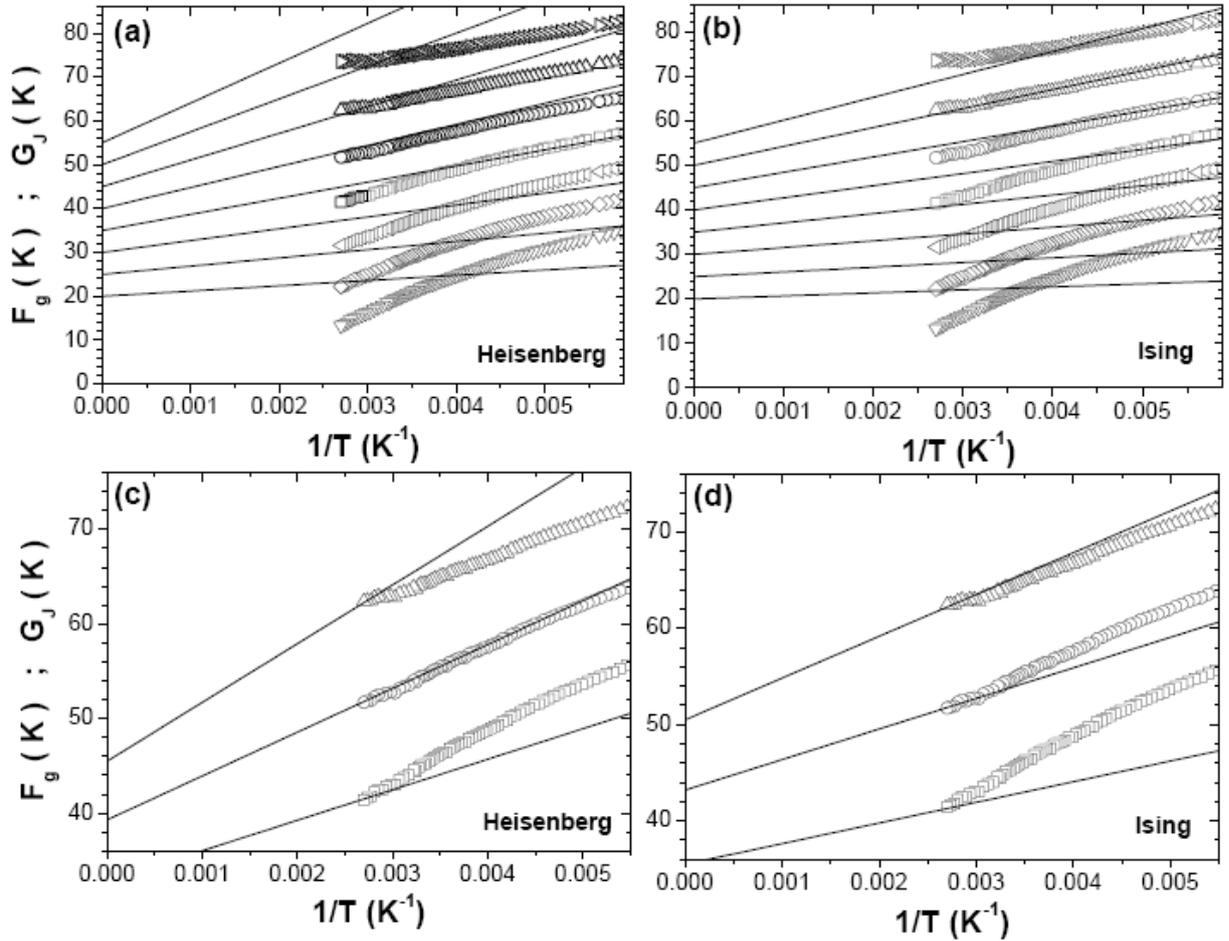

**Fig. 9** Analyses of the high-T regime via the formula of series expansions calculated for Heisenberg or Ising models, with S=3/2. Each panel displays the function $F_g(\chi,T)$ calculated from the experimental $\chi(T)$ and for various values of the Landé factor g (open symbols), along with plots of $G_J(T)$ (solid lines) for various values of the intrachain coupling J (in Kelvin) and α corresponding either to the Heisenberg model (α = 3) or the Ising model (α = 17/10). The J values of the solid lines simply correspond to the intercepts with the vertical axis. The different g values are associated with the following symbols: g=2 (triangles down); g=2.1 (diamonds); g=2.2 (triangles left); g=2.3 (squares); g=2.4 (circles); g=2.5 (triangles up); g=2.6 (triangles right).The panels (a) and (b) display the same sets of *J* and g values. The panels (c) and (d) focus on the interval of g values leading to linear regimes of best quality in the range T > 200 K. In these panels, we selected *J* values such as the solid lines match the data at the highest temperature.

The high-T regime considered for our calculations is about T > 200 K. Fig. 9 (a) and (b) show $F_g[\chi(T),T]$ for a series of g values as well as $G_J[T]$ for a series of *J* values (Heisenberg in Fig. 9 (a) and Ising in Fig. 9 (b)). We emphasise that $F_g[\chi(T),T]$ must be linear especially in the highest T range, i.e. the left part of the plots of Fig. 9. For g = 2, one observes that the extrapolation of the linear part of $F_g[\chi(T),T]$ leads to positive values for *J*, which are not physically valid, whereas, for g = 2.6, an upturn develops in $F_g[\chi(T),T]$ at the highest temperatures. Actually, a reasonable range of g values is between 2.3 and 2.5. This is the range shown in Fig. 9 (c) and 9 (d). For each g value, $G_J[T]$ is plotted for the *J* value matching the data at the highest temperature, in order to compare the Heisenberg and Ising limit in a reliable manner (solid lines in Fig. 9 (c) and 9 (d)). The best overlap between $F_g[\chi(T),T]$ and $G_J[T]$ is obtained within the framework of the Heisenberg limit, for g $\approx$ 2.4 and *J* $\approx$ -39 K. It is worth noticing that this behaviour is consistent with the literature, in that $Co^{2+}$ in such distorted tetrahedral environment generally leads to almost isotropic g values lying in the range 2.2-2.4.[35] Moreover, Palii *et al*.[30] have previously noted that the Heisenberg form of the exchange coupling has been successfully used in description of the numerous $Co^{2+}$ cluster compounds.

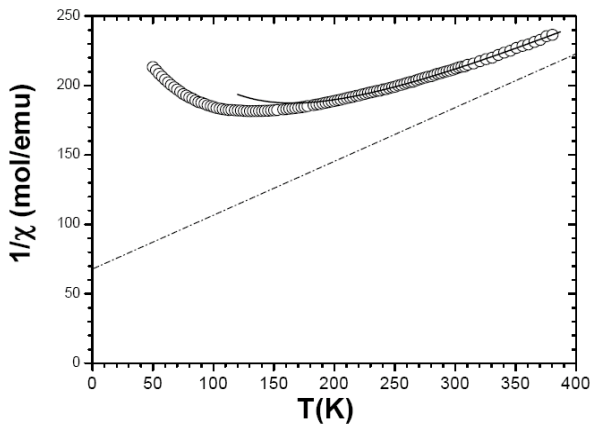

**Fig. 10** Reciprocal susceptibility for T > $T_N$ (circles); Results of the series expansion (Heisenberg) for *J* = -35.1 K and *g* = 2.35 (solid line); Curie-Weiss plot corresponding to the same values of *J* and *g*. The dotted line represents the C-W model.

In a second step, we refined the values of the parameters $J$ and g by considering the first six terms in the series expansion [COEFF] and fitting the data in the high-temperature range. The coefficients $b_n$ of the series expansion for S= 3/2 Heisenberg spin chains were calculated from the formula derived by Rushbrooke and Wood.[34] We found: $b_1$ = -5; $b_2$ = +15; $b_3$ = -21.333; $b_4$ = -18.542; $b_5$ = +118.075; $b_6$ = +16.035. Using these parameters, the refined values for the intra-chain coupling and the Landé factor are $J \approx$ -35.1 K and g $\approx$ 2.35, respectively.

Fig. 10 shows a comparison between the experimental curve and the curve calculated using the series expansion with the above parameters. There is a good fitting at high temperature, and this fit well accounts for the rounding of the experimental curve. On the other hand, the calculated curve departs from the data at lower temperatures, as expected. In Fig. 10, we also plotted the CW function from Equations (2) and (3), for $J \approx$ -35.1 K and g $\approx$ 2.35. A substantial shift is found between the CW plot and our curve, demonstrating that the CW regime can just be regarded as the limit towards which the data tend at very high temperatures.

### 3.3.2  Intermediate-T range of the χ(T) curve.

We adopted alternative approaches to calculate the value of $J$, using models specific to one-dimensional magnetic systems. Consistently with the above conclusion of the series expansion analysis, we only considered models corresponding to the Heisenberg limit.

**3.3.2.1 Numerical results on finite chains.** For spin chains, the coordinates of the broad maximum in χ(T), i.e. $T_{max}$ and $\chi_{max}$, are related to the value of $J$. Within the Heisenberg limit, exact results are available for S=1/2[36] or for S tending to infinity.[37] However, numerical results were also obtained for intermediate S values by considering finite chains. Most of these results were compiled by De Jongh and Miedema in their review paper.[3] For Heisenberg antiferromagnetic chains with S=3/2, the following values are predicted:

$$\frac{T_{max}}{|J|} \approx 4.75 \qquad (8a)$$

$$\frac{k_B |J| \chi_{max}}{N_A g^2 \mu_B^2} \approx 0.091 \qquad (8b)$$

In our case, $T_{max} \approx$ 135 K and $\chi_{max} \approx$ 0.0055 emu/mol. From Equation 5 a, $J \approx$ -28.4 K. Equation 8 b requires a value for the Landé factor. By substituting g $\approx$ 2.35 in Equation (8 b), $J \approx$ -34.2 K is obtained.

**3.3.2.2 Wagner-Friedberg model.** As noted above, none of the models known to date can fit the T dependence of the susceptibility over the whole T range for a Heisenberg antiferromagnetic spin chain with S=3/2. However, Fisher's model 37 can be modified to account for the case of finite S values different from 1/2. This was done by Wagner and Friedberg who proposed the following formula:[38]

$$\chi(T) = \left(\frac{N\mu_B^2}{k_B}\right)\left(\frac{g^2 S(S+1)}{3}\right)\frac{1}{T}\frac{1 - u(T_0/T)}{1 + u(T_0/T)} \qquad (9)$$

with

$T_0 = 2|J|S(S+1)$

$u(x) = \coth(x) - 1/x$

This function was already used with success for different spin chains.[6,11,38] It must be noted ed that Equation 6 is not expected to fit the data within the whole T range but only to account for data at temperatures T > $T_{max}$. Fig. 11 shows that the Wagner – Friedberg model gives a good fit of the data for T > $T_{max}$, but cannot reproduce the drop in χ for T < $T_{max}$. By fitting to the data over the range 110-380 K, we derived the following set of parameters: $J \approx -38.6$ K and $g \approx 2.37$.

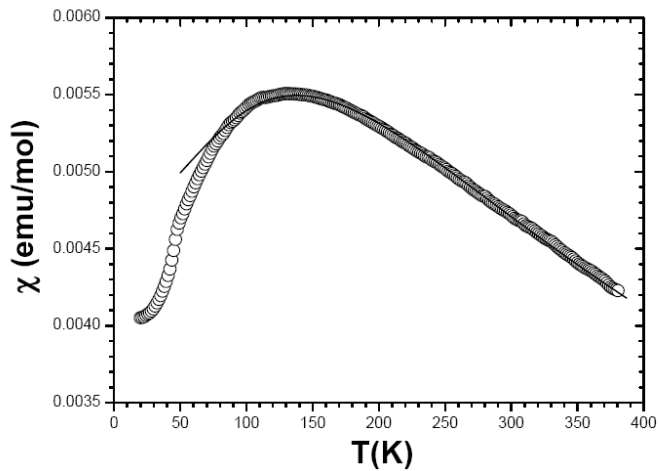

**Fig. 11** Susceptibility curve in an expanded scale, along with the fitting curve of the S=3/2 Heisenberg model (Wagner and Friedberg), for $J = -38.6$ K and g = 2.37.

**3.3.2.3 Emori model.** In the previous investigation of $Ba_2CoS_3$,[11] the χ (T) curve was analysed with the reduced-spin model proposed by Emori et al.[9] This model is an empirical extension of the S=1/2 Ising model, in which the actual values of S and J are replaced by an effective spin S* = √[S(S+1)/3] and by the corresponding effective coupling $J^*$. The $\chi$ (T) at high-T is fitted to the following expression:

$$\chi = \left(\frac{N_A \mu_B^2}{k_B}\right)\left(\frac{g^2}{3}\right)\frac{S(S+1)}{T}\exp\left(\frac{J^*}{T}\right) \quad (10)$$

The relationship between the true coupling, $J$, and the effective one, $J^*$, is $J = 3J^*/4S(S+1)$.

From a linear regime in a semi-log plot of $3\chi T/[N_A\mu_B^2/k_B]$ versus 1/T, g and $J$ can be derived. Nakayama et al.11 showed that such a regime is obeyed in $Ba_2CoS_3$. Our data confirm this, showing a linear regime for T > 200 K. From this fitting, $J \approx$ -35.7 K for g = 2.34 can be derived. It must be pointed out that our J value is considerably larger than that reported by Nakayama ($J \approx$ -12 K).11

At first glance, it is puzzling that the Emori model, claimed to deal with Ising-like systems, can fit the susceptibility data for $Ba_2CoS_3$, while we have shown that this compound is more of a Heisenberg nature. We suggest that the definition "Ising-like" may be not appropriate for the Emori model, which should instead be regarded as an averaging method, where the partition function is approximated by only two terms corresponding to $-S^*$ and $+S^*$, instead of all the 2S+1 projections associated with the true spin S.

Moreover, previous studies have shown that the Emori model can fit the susceptibility data for $Ba_2MnS_3$, even though this compound is recognised to be a Heisenberg system, as $Mn^{2+}$ is a spin-only cation almost insensitive to the CEF.[6,11] We prepared a ceramic sample of f $Ba_2MnS_3$ and measured its susceptibility versus temperature. The Emori model was used to fit data over the high temperature range in the plot of the reciprocal susceptibility versus temperature. A good fit was obtained, leading to a value for $J$ ($\approx$ -14.8 K) that is close to the one calculated using the Wagner-Friedberg model ($\approx$ -12 K).[6,10]. These results show that the e definition "Ising-like" is not appropriate for the Emori model.

### 3.3.3 Comparison of the models

In Fig. 12, all the calculated patterns fitting the susceptibility data are compared in a plot of 1/$\chi$ versus T.

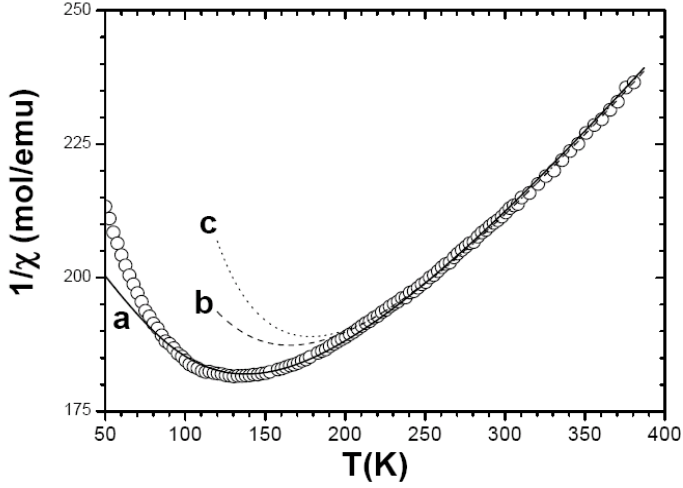

**Fig. 12** Enlargement of the reciprocal susceptibility for T > $T_N$, comparing the quality of the fitting curves derived from three different approaches: (a) the Wagner-Friedberg model (b) the Rushbrooke-Wood model; (c) the Emori model

Fig. 12 shows that the Wagner-Friedberg model provides the best overall fitting, down to about $T_{max}$. Moreover, it shows that the Emori model and the series expansion lead to fitting curves of similar quality but in a narrower temperature range.

Let us now combine the results of these various approaches in order to get the best estimate of $J$. The case of $Ba_2MnS_3$ which has been more widely investigated can help to evaluate the reliability of the models discussed in this paper. We followed analogous procedures to calculate $J$ for $Ba_2MnS_3$. The values we obtained are: $J \approx -11$ K (from $T_{max}$); $J \approx -14.9$ K (from $\chi_{max}$); $J \approx -16.0$ K (from the series expansion); $J \approx -15.0$ K (from the Wagner-Friedberg model); $J \approx -14.8$ K (from the Emori's model). These results indicate that the formula derived from numerical calculations in finite chains, particularly the one based on the $T_{max}$ value tends to underestimate $J$. We note that this feature also appears in the data gathered by De Jongh and Miedema[3], for instance in the case of S=5/2 compounds such as $CsMnCl_3·2H_2O$ and $[(CH_3)_4N][MnCl_3]$ for which the $J$ determined from $T_{max}$ are about 20% smaller than the ones obtained from the other models.

Therefore, in order to present average experimental values for $J$ and g in $Ba_2CoS_3$, we prefer to restrict ourselves to the results of the series expansion and the Wagner-Friedberg model. This leads to $J = -37 \pm 2$ K with g = 2.36 ± 0.01.

### 3.4   Three-dimensional long-range ordering at $T_N$

### 3.4.1. Signature of $T_N$ on $\chi(T)$.
Measurements of $\chi(T)$ were recorded in a magnetic field of 1 T, in order to obtain better resolved data than the ones shown in previous studies.[4,11] The main panel of Fig. 13 shows an enlargement of the low-T part of the $\chi(T)$ curve.

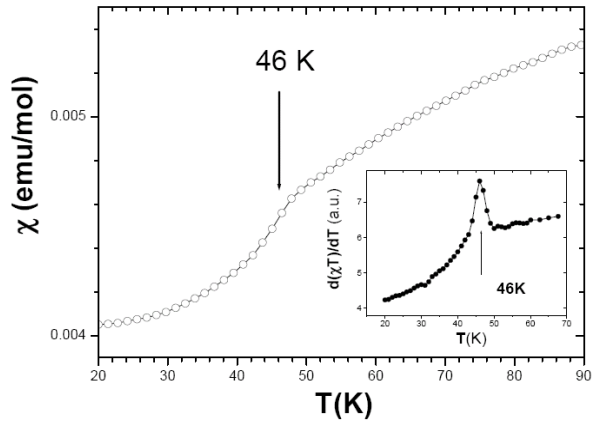

**Fig. 13** Main panel : Temperature dependence of the magnetic susceptibility. The arrow indicates a change of regime, corresponding to a steeper drop in $\chi$ as T is decreased. Inset: Temperature dependence of the derivative of $\chi T$, showing a peak at 46 K

One can see that, below a temperature T ~ 46 K there is a change in the temperature dependence of $\chi$ (marked by an arrow). This change of regime is still more visible in the inset of Fig. 13, which shows the derivative of $\chi T$ versus the temperature. On this latter plot, one observes a prominent peak located at 46 K. Since the susceptibility $\chi$ decreases rapidly below 46 K, one can speculate it marks the onset of a three-dimensional, long-range antiferromagnetic ordering within the spin system. Supporting this picture is the fact that the overall shape of our $\chi(T)$ closely resembles curves typical of spin chains systems displaying the onset of a 3D long-range spin ordering at a Néel temperature ,$T_N$, resulting from the combination of antiferromagnetic intra- and inter-chain couplings.[39,40,41]

To investigate further the possibility of a Néel temperature at 46 K, we measured the heat capacity of $Ba_2CoS_3$ as a function of temperature.

### 3.4.2 Signature of $T_N$ on C(T).
The heat capacity is known to be a powerful method to detect the onset of a long-range magnetic order.

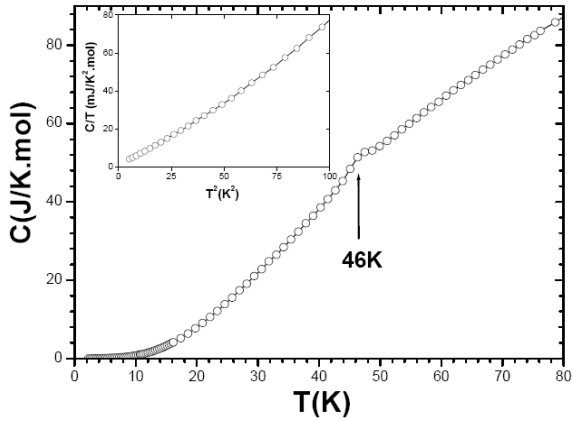

**Fig. 14**. Main panel: temperature dependence of the total heat capacity (lattice and magnetic contributions), showing a peak at 46 K. Inset: Low-T range in a C/T vs. $T^2$ plot.

Fig. 14 shows the dependence of the heat capacity, C, with the temperature, T, in zero magnetic field. Here, we focus on the range of temperature 2-80 K, as no anomaly was detected at higher temperatures. On the C(T) curve, a small peak appears at 46 K, an indication of a long-range transition at this temperature. It must be specified that the peak is not very pronounced since C(T) includes, among several contributions, a phononic term which is predominant in this temperature range. Accordingly, a "magnetic peak" associated with $T_N$ is thus expected to be quite small on the curve of the total heat capacity, as in the curve in Fig. 14.

Furthermore, one observes that the peak in the C(T) curve appears exactly at the same temperature as the peak in $d(\chi T)/dT$ (Fig. 13, inset). It must be emphasised that this feature is theoretically expected for an antiferromagnetic long-range ordering,3 a result which strongly supports the fact that the transition at 46 K is actually a $T_N$. It can also be noted that such an antiferromagnetic nature of the 3D spin transition is consistent with a Mössbauer study which evidenced antiferromagnetic order at 4.2 K. 11.

The inset of Fig. 14 shows a *C/T* versus $T^2$ plot in the temperature range 0 - 10 K. The curve appears as a straight line with a parabolic-like trend at higher temperature. The heat capacity is the sum of different contributions. In the simplest situation, *C(T)* is dominated by the electronic and phononic terms, leading to $C = aT + bT^3$ where $aT$ is the electronic term and $bT^3$ the phononic term. In our case, a magnetic term—due to spin-wave excitations in the ordered spin system—should be added. In the case of three-dimensional antiferromagnetic order, this term is expected to be cubic in temperature, i.e., of the form $dT^3$ and, even when including this magnetic term, a $C/T$ vs. $T^2$ plot is still expected to show a linear behaviour. At higher temperatures, though, the phononic term progressively deviates from the pure cubic behaviour and a correction term of the form $cT^5$ should be included, which leads to $C = aT + bT^3 + dT^3 + cT^5$. In this case the $C/T$ vs $T^2$ plot shows linear behaviour, with a parabolic-like curvature appearing at the highest temperatures. This is precisely the behaviour showed in the inset of Fig. 14.

### 3.5 Estimation of the inter-chain coupling $J'$

In a one-dimensional system, the temperature of long-range magnetic ordering is a combination of the intra- and inter-chain couplings, referred to as $J$ and $J'$, respectively. For Heisenberg systems with antiferromagnetic intra- and inter-chain coupling, Oguchi has derived the following relationship from the Green function method:[42]

$$T_N = \frac{4S(S+1)|J|}{3} \frac{1}{I(\eta)} \text{ with } \eta = |J'/J| \qquad (11)$$

where $\eta$ is the ratio between the inter- and intra- chain coupling. This relation assumes that the chains are arranged on a square lattice, which is actually the case for $Ba_2CoS_3$. In the paper by Oguchi et al., the integral $I(\eta)$ is tabulated for a series of $\eta$ values. An interpolation function was proposed by Ami et al.: $I(\eta) = 0.64/\sqrt{\eta}$.[43]

Including this expression of $I(\eta)$ in Equation 8, one derives

$$J' = \frac{1}{J}\left[\frac{1.92\, T_N}{4S(S+1)}\right]^2 . \qquad (12)$$

Using this equation with $T_N$ = 46 K, $J$ = -37 K and $S$ = 3/2, leads to $J' \approx$ - 0.9 K.

Once the order of magnitude of $J'$ is determined, it is important to evaluate the influence of this inter-chain coupling on the raw $\chi$ (T) data. So far, the susceptibility has been analysed within the one-dimensional scenario i.e. neglecting the inter-chain coupling. Within a mean-field approach, the pure free-chain susceptibility $\chi_{1d}$ (T) is linked to the raw data $\chi$ (T) by the relationship:[44]

$$\chi_{1d} \approx \frac{\chi}{1 + \frac{2z'J'}{N_A(g\mu_B)^2}\chi} \qquad (13)$$

In the $Ba_2CoS_3$ case, $z'$ = 4 since each chain has four nearest neighbours. Using Equation (10) with $J'$ = - 0.9 K, and $g$ = 2.36, it is found that $\chi_{1d}$ (T) remains very close to $\chi$ (T).

Calculations carried out using $\chi_{1d}$ (T) instead of $\chi$ (T), in all the models considered in this paper, showed that the value of $J$ is not substantially modified.

## 4. Conclusions

Recently, we had found that $Ba_2CoS_3$ shows peculiar transport properties, including metallic-like behaviour and small negative magnetoresistance. In the present paper, we report on the magnetic properties of this compound.

To start with, we investigated the local structure of the cobalt ions in $Ba_2CoS_3$, by using EXAFS and XANES. This study confirmed the +2 oxidation state and provided a detailed characterisation of the distorted tetrahedral coordination of $Co^{2+}$ in $Ba_2CoS_3$.

On these grounds, we performed ab-initio calculations, which demonstrated that the $Co^{2+}$ are in a high-spin state (S=3/2). Using a perturbative approach, it was also shown that the intra-chain coupling between two adjacent $Co^{2+}$ via a $S^{2-}$ must to be antiferromagnetic.

We proceeded then to derive this antiferromagnetic intra-chain coupling J from the temperature dependence of the magnetic susceptibility, using and comparing several models. We found $J$ = -37 ± 2 K with a Landé factor g = 2.36 ± 0.01. Finally, we identified the existence of a three-dimensional (3D) magnetic ordering, in both the susceptibility and heat capacity data. This transition —which occurs at 46 K— is shown to be a Néel temperature, marking the onset of a 3D long-range antiferromagnetic ordering. We estimated the inter-chain antiferromagnetic coupling $J' \approx$ -1 K. The ratio $J'/J$ shows that $Ba_2CoS_3$ should be regarded as a *quasi 1-D* system.

## 5. Appendix

The coupling matrix between two neighbouring $Co^{2+}$ ions can be expressed in the total $S_Z$=0 sector as displayed in Table 3. The basis set supporting this coupling matrix is as the following, where *l* and *r* exponents refer to the left and right atoms, the barred orbitals are assumed to be $S_Z$=-1/2 and the non-barred ones $S_Z$=1/2.

The formal diagonalisation of this matrix can be carried out and assuming that all the elementary exchange integrals are antiferromagnetic, that is negative in the present convention, the ground state can be easily determined. Its energy is

$$E_{GS} = J_c + J_{yz} + \frac{J_{xz}}{2} + \frac{J_{x^2-y^2}}{2} - \frac{1}{2}\sqrt{(J_{x^2-y^2} + J_{xz} - J_c)^2 + 3J_c^2}$$

and the wave function is

**ACKNOWLEDGEMENTS**

The authors thank the EPSRC and the Pierre and Marie Curie fellowship for financial support. Thanks also to CCLRC for access to synchrotron and computing facilities at Daresbury Laboratories. The authors also thank Dr. D. Maynau for providing the CASDI suite of programs.